\documentclass[a4paper,12pt,twocolumn]{article}
\usepackage{cite}
\usepackage[dvips,hiresbb]{graphicx}
\renewcommand{\figurename}{Fig.}

\title{\textbf{Special polarization characteristic features of a three-dimensional terahertz photonic crystal quantitatively evaluated by phase-differences with a FEM}}
\author{\Large{Chikara Sakurai} \thanks{Email: c-sakurai@river-ele.co.jp (first), sakuraikazan@gmail.com (second)} \\ \textit{River Electric Corporation, 2-1-11 Fujimigaoka, Nirasaki, Yamanashi, Japan}} 
\date{\today}
\begin{document}
\onecolumn{
\maketitle
\begin{abstract}
\normalsize{A previous work, ref.~\cite{sakurai2} experimentally confirmed that measured special polarization characteristic features of a three-dimensional terahertz (THz) photonic-crystal  did not apparently apply to general physical- and optical- basic rules. It is said that three-dimensional (3D) photonic crystals (PC) have no polarization anisotropy within photonic band gap (PBG, stop gap, stop band) of high symmetry points in normal incidence. None the less because two directions at right angles to each other on the plane (001) are constitutionally and optically identical \textit{in appearance}, the polarization orientation of a reflected wave rotated by 90 degrees from that of an incident wave at a frequency within the PBG when that of the incident wave was the direction that bisects these two directions.
The 3D-PC is a silicon inverse diamond structure whose lattice point shape is vacant regular octahedrons.
In this work, the phase-difference, size and orientation of electric-field of the reflected wave, had been studied by using a FEM (finite element method). The analyses indicated the reason why the polarization orientation of the reflected wave is rotated by 90 degrees at the frequency within the PBG. In addition, it was found that the orbit of electric-field of the reflected wave becomes near-circularly-polarized one at another frequency within the PBG for the linear-polarization incident wave.} 
\end{abstract}
\twocolumn
\section*{Introduction}
\hspace*{5mm}Special polarization characteristic features of a three-dimensional (3D) terahertz photonic-crystal (PC) have been studied in previous works, ref.~\cite{sakurai1}, \cite{sakurai2} and \cite{sakurai3} \footnote{ Ref.~\cite{sakurai2} contains more detailed analyses than those in ref.~\cite{sakurai3}.}. These features on the polarization anisotropy are those related to the polarization orientation (electric-field direction) difference between  reflected wave and incident one.\\
\hspace*{5mm}The 3D-PC is a silicon inverse diamond structure whose lattice point shape is vacant regular octahedrons. Measured polarization anisotropy$^{[2], [3]}$ in normal incidence is that the polarization orientation of a reflected wave rotates by 90 degrees
to a polarization orientation of an incident wave at a frequency within BGX that is X point's photonic-band-gap (PBG).\\
\hspace*{5mm}Fig.~\ref{fig:2polar}$^{[1]}$ shows the comparison of the polarization characteristic features as generally expected with ones in the measured spectra.\\
\hspace*{5mm}These results did not apparently apply to general physical- and optical- basic rules as follows.\\
\hspace*{5mm}(A) Electric-field is vector, and various optical phenomena are analyzed by using methods of resolution and synthesis of it.\\
\hspace*{5mm}(B) According to Maxwell's equations, 
the polarization orientation of the reflected wave rotates by 180 degrees from that of the incident wave in the case of complete reflection (Bragg reflection) and normal incidence.\\
\hspace*{5mm}(C) According to crystallography, diamond structure is in cubic system and optically isotropic.\\
\begin{figure}[t]
\centering
\includegraphics[width=7.8cm,]{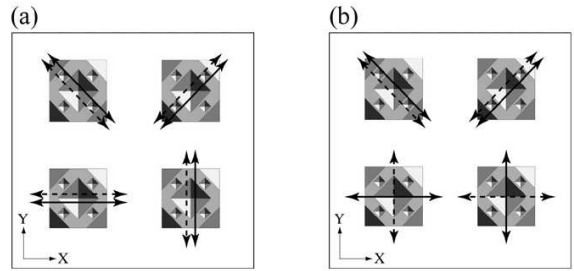}
\caption{\footnotesize (a) Features expected from basic and optical rules. (b) Special characteristic features in this work. The incident wave is normal incidence, [001]. The heavy solid arrow is the polarization orientation (electric-field direction) of the incident wave, and the heavy dashed arrow is that of the reflected wave. The unit cell of a diamond structure in fig.~\ref{fig:latticeA} is illustrated.}
\label{fig:2polar}
\end{figure}
\hspace*{5mm}Polarization features in ref.~\cite{min} do not also apply to rule (B) and (C).
The features, however, exist in the limit of low-frequency wave; the wavelength is much larger than the lattice constant of photonic crystal. In addition, the frequency is without PBG, and then the 3D-PC has eigen modes (branches). An eigen mode have an intrinsic symmetry, and some polarization anisotropy also exists. On practical side, the reflectivity is small.\\
\hspace*{5mm}In previous and this works, however, the wavelength of the incident wave is comparable to the lattice constant, and the frequency is within BGX. No eigen modes based on the 3D-PC band structure exist within BGX.\\
\hspace*{5mm}None the less because some two directions at right angles to each other on the plane (001) are constitutionally and optically identical \textit{in appearance}, the polarization orientation of the reflected wave rotated by 90 degrees from that of the incident wave at a frequency within BGX when that of the incident wave was the direction that bisects these two directions. These facts do not apparently apply to rule (A).\\ 
\hspace*{5mm}In ref.~\cite{sakurai1}, measured reflection spectra were semi-quantitatively evaluated by being compared with reflection ones analyzed by a
FEM (finite element method). These results suggest that measured polarization anisotropy does not apply to Maxwell’s equations \textit{in appearance} but
applies to these equations in essential as expected.\\
\hspace*{5mm}In this work, the phase-difference, size and orientation of electric-field of two reflected waves whose polarization orientations are identical \textit{in appearance}, have been studied primarily and quantitatively for resolution of polarization rotation.\\
\hspace*{5mm}In addition, it was found that the phase-difference gradually shifts when the frequency shifts within BGX, and then the reflected wave becomes near-circularly-polarized wave at another frequency within BGX.\\
\section*{Parameters}
\hspace*{5mm}The unit cell of the simulation model is the diamond structure as shown 
in fig.~\ref{fig:latticeA}(a). The sphere is the lattice point and its shape is the regular octahedrons in fig.~\ref{fig:latticeA}(b). 
It is vacant (atmosphere) and the dielectric constant, $\varepsilon_1$ = 1.00 is set. 
The surrounding material is pure Si and the dielectric constant, $\varepsilon_2$ = 11.9 is set.
The lattice constant, $a$ = 300 $\mu$m and the side length of the regular octahedrons, $L$ = 150 $\mu$m are set in the theoretical and experimental works$^{[1]}$.
In the unit cell of the Si inverse diamond structure, many sections of the air octahedrons on the lattice points are arranged as shown in fig.~\ref{fig:latticeA}(c) and \ref{fig:latticeA}(d). \\    
\hspace*{5mm}Fig.~\ref{fig:bandstructure}(a) shows the calculated photonic band structure~$^{[3]}$ by using plane wave expansion method.
In a series of works, the direction of the incident wave is +Z-direction, [001]
in the real space and it corresponds to $\Gamma$-X direction in the wave number space (K-space). 
The polarization anisotropy is studied on the surface (001)\footnote{[001] and (001) are defined on the X-Y-Z coordinate system in fig.~\ref{fig:latticeA}.} at around BGX. \\
\begin{figure}[t]
\centering
\includegraphics[width=6.5cm,]{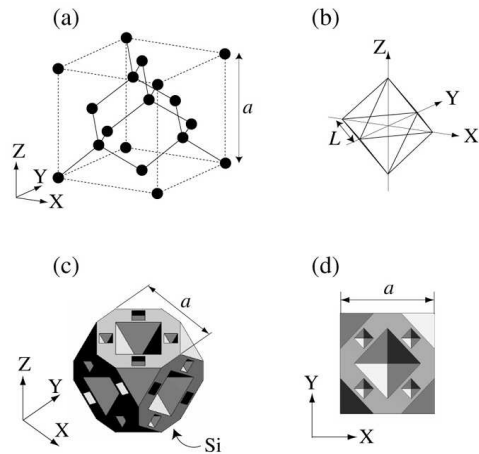}
\caption{\footnotesize (a) Lattice of the diamond structure. (b) Shape of the lattice point is regular octahedrons. It is vacant. The surrounding material is Si. (c), (d) Unit cell of the Si inverse diamond structure.}
\label{fig:latticeA}
\end{figure}
\begin{figure}[t]
\centering
\includegraphics[width=7.4cm, trim=0 0 0 0]{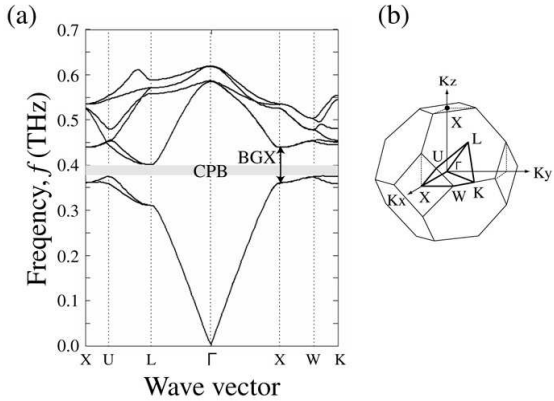}
\caption{\footnotesize (a) Calculated photonic band structure has complete photonic band gap, CPB (gray zone) at around 0.4 THz. BGX exists between 0.36 THz and 0.44 THz. (b)First Brillouin zone and the reduced zone (heavy line) with high symmetry points.}
\label{fig:bandstructure}
\end{figure}
\hspace*{5mm}Fig.~\ref{fig:IXY} shows the definition of polarization orientation (electric-field direction) of the incident wave, PI(XY). The unit cell of the diamond structure in fig.~\ref{fig:latticeA} is illustrated.
For example, PI(1, 0) and PI($-$1, 0) are indistinguishable; they are collectively expressed as PI(10).
Furthermore, PI(11) is parallel to function, $Y=X$. PI(10), PI(01), PI(11) and PI(1$\overline{1}$) are expressed as [X], [Y], [Y = X] and [Y = $-$X], respectively \footnote{PI(X,Y) corresponds to I(X,Y) in ref.[1]. PI(11) and PI(1$\overline{1}$), however, need to be differentiated in this work.}. The latter is also used as the indication of polarization orientation of the reflected wave. \\
\hspace*{5mm}Four analyzed models are shown in fig.~\ref{fig:modelAB}. X-Y-Z coordinate system corresponds to that in fig.~\ref{fig:latticeA}.
The model size is $L_\mathrm{a}$ $\times$ $L_\mathrm{a}$ $\times$ $L_\mathrm{b}$ ($\mu$m)$^3$. $L_\mathrm{b}$ is the height of the Z-direction. $L_\mathrm{a}$ $\times$
$L_\mathrm{a}$ is the area in X-Y plane. 
The square area surrounded with white lines is $a$ $\times$ $a$ ($\mu$m)$^2$, which corresponds to the unit cell in fig.~\ref{fig:latticeA}(d).
\begin{figure}[t]
\centering
\includegraphics[width=4cm, trim=0 0 0 0]{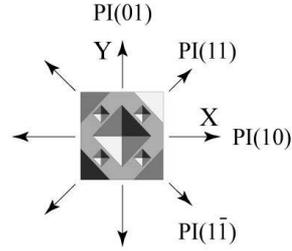}
\caption{\footnotesize Four kinds of PI(XY) is the polarization orientation of the incident wave in normal incidence. }
\label{fig:IXY}
\end{figure}
The area of the incident wave, port is shown as the square surrounded with black heavy lines, whose size is $L_\mathrm{p}$ $\times$ $L_\mathrm{p}$ ($\mu$m)$^2$.
The white heavy arrows are the orientation of electric-field vector of the incident wave,
which is fixed. The direction of the incident wave is +Z-direction and normal incidence. Each model is rotated instead of the electric-field vector of the incident wave. $(L_\mathrm{a}, L_\mathrm{b}, L_\mathrm{p})$ are set as $(6a, 5a, 5a) (\mu$m). 
In model(11), model(1$\overline{1}$), model(10), and model(01), the electric-field vector is parallel to [Y = X], [Y = $-$X], [X], and [Y], respectively.\\
\hspace*{5mm}The maximum value of side of the model size is 1800 $\mu$m.  The models were analyzed with the size magnified a hundred times since the maximum drawing size in a FEM software is 1000 $\mu$m and 1000 mm when analysis unit is $\mu$m and mm, respectively. The scaling rule was applicable in the photonic crystals. In other words, frequency $\times$ size (lattice constant) is constant. On the base of this rule, the spectrum frequencies were finally shifted from GHz to 0.1 $\times$ THz as the size needed to be reduced by a hundredth. The following explanations use final sizes ($\mu$m) and final frequencies (THz). Models' mesh size is 50 $\mu$m\footnote{The mesh size is nearly minimum for~the limit of the personal computer's performance: OS: Windows~10~Home (64bit), processor: Intel(R) Core (TM) i5-6500 CPU@3.20 GHz, RAM: 64.0 GHz.}.
\begin{figure}[t]
\centering
\includegraphics[width=7.5cm, trim=0 0 0 -15]{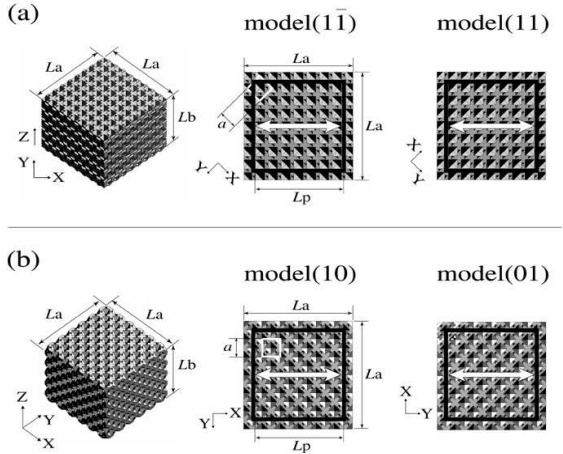}
\caption{\footnotesize In four analyzed models, the square surrounded with black heavy lines, port is the area of the incident wave. Each white heavy arrow is the orientation of electric-field vector of the incident wave, which is fixed. Instead, the models are rotated. }
\label{fig:modelAB}
\end{figure}
\section*{Simulation Results}
\hspace*{5mm}In this work, the reflected spectra and these phases were primarily analyzed by using a FEM (finite element method)\footnote{The FEM software was Femtet(R)2017.1 made in Murata Software Co., Ltd..}.
The FEM software had no command to obtain S-polarization (S-p) spectra and P-polarization (P-p) ones, separately. Only compound spectra of S-p and P-p \footnote{Two polarization are explained in detail in fig.~A.\ref{fig:RTsystem}
(appendix).} were able to be obtained.\\
\hspace*{5mm}These spectra in figs.~\ref{fig:RT4} (a) to \ref{fig:RT4}(d) are obtained by using model(1$\overline{1}$), model (11), model(10) and model(01) in fig.~\ref{fig:modelAB}, respectively. The reflected spectra (closed circle) and transmission ones (open circle) are shown. The variable, $f$ is THz-frequency.\\
\begin{figure}[t]
\centering
\includegraphics[width=7.4cm, trim=10 0 0 0]{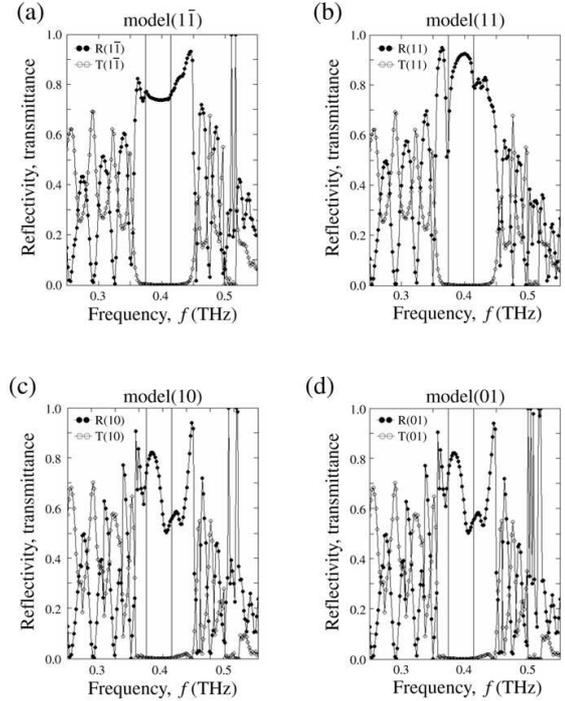}
\caption{\footnotesize Reflected spectra (closed circle), R(XY) and transmission ones (open circle), T(XY) in each model(XY) in fig.~\ref{fig:modelAB}.}
\label{fig:RT4}
\end{figure}
\hspace*{5mm}Four S-parameters are calculated by using the software, which are $S_{11}$,  $S_{22}$, $S_{21}$ and $S_{12}$. The reflectivity and the transmittance correspond to $|S_{11}|^2$ and  $|S_{21}|^2$, respectively. The analyzed frequency range is 0.25 THz to 0.55 THz, in which BGX (0.36 THz to 0.44 THz) is included. The frequency interval of the data is 0.002 THz. The absorption coefficient (cm$^-$$^1$) is set as zero. Two vertical solid lines are ones at 0.374THz and 0.414THz, which are explained in fig.~\ref{fig:phaseB}.\\
\hspace*{5mm}These spectra are S-p and P-p compound ones as stated above. 
The orientation of S-p is parallel to that of polarization (electric-field) of the incident wave, PI(XY). PI(XY) is parallel to the white heavy allow in fig.~\ref{fig:modelAB}. The orientation of P-p is perpendicular to that of S-p. The symbol, (XY) of the reflectivity, R(XY) and transmittance, T(XY) corresponds to that of PI(XY).\\
\hspace*{5mm}PBG's in figs.~\ref{fig:RT4} (a) to \ref{fig:RT4}(d) are nearly equal to BGX (0.36 THz to 0.44 THz). The transmittance in each model is very small within PBG.  In four reflected spectra within PBG, the spectrum shape in model(1$\overline{1}$) displays convex-downward one, and the shapes in other models do concavity and convexity.
The shape in model(10) is nearly equal to that in model(01) within PBG.\\
\hspace*{5mm}
In the sub-peaks of reflected spectra and transmission ones without PBG, local maximum value of the former and local minimum one of the latter tend to appear at the same frequency as expected.\\
\hspace*{5mm}The phases of four reflected spectra are shown in fig.~\ref{fig:phaseA}. 
$\theta$(XY) is the phase of R(XY), which is the phase-difference between the incident wave and reflected one, in model(XY) in fig.~\ref{fig:modelAB}.
Fig.~\ref{fig:phaseA}(a) shows $\theta$(11) (closed circle) and $\theta$(1$\overline{1}$) (open circle). $\theta$(10) (closed circle) and $\theta$(01) (open circle) are shown in fig.~\ref{fig:phaseA}(b).
$\theta$(10) and $\theta$(01) have almost the same phase-frequency features.
On the other hand, $\theta$(11) and $\theta$(1$\overline{1}$) have entirely different phase-frequency ones especially within PBG. A frequency at which the phase-difference between $\theta$(11) and $\theta$(1$\overline{1}$) is 180 degrees or 90 degrees, exists.\\
\begin{figure}[h]
\centering
\includegraphics[width=7.4cm, trim=10 0 0 0]{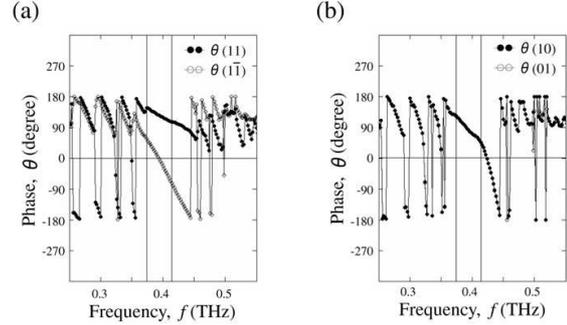}
\caption{\footnotesize $\theta$(XY) is the phase of R(XY). It is the phase-difference between the incident wave and reflected one.}
\label{fig:phaseA}
\end{figure}
\hspace*{5mm}The phase-difference, $\theta(11)-\theta(1\overline{1})$ is shown as $\Delta\theta_{A}$ in fig.~\ref{fig:phaseB}(a).
In fig.~\ref{fig:phaseB}(b), $\Delta\theta_{B}$ is the phase-difference, $\theta(10)-\theta(01)$.
$\Delta\theta_{A}$ is 90.7 degrees ($\doteq$ 90 degrees) at 0.374 THz, and then is 179.4 degrees ($\doteq$ 180 degrees) at 0.414 THz, in the frequency interval, 0.002 THz. These two frequencies are shown as vertical solid lines.\\
\hspace*{5mm}Compared with $\Delta\theta_{A}$ without PBG, $\Delta\theta_{A}$ within PBG  gradually shifts from 90 degrees to 180 degrees. On the other hand, $\Delta\theta_{B}$ within PBG is nearly equal to zero.\\
\hspace*{5mm}These phase-difference features are, naturally, important to explain polarization features at above two frequencies within PBG. Furthermore, the information of orientation and size of electric-field of the reflected-wave is also needed to the same degree. \\
\begin{figure}[t]
\centering
\includegraphics[width=7.4cm, trim=10 0 0 0]{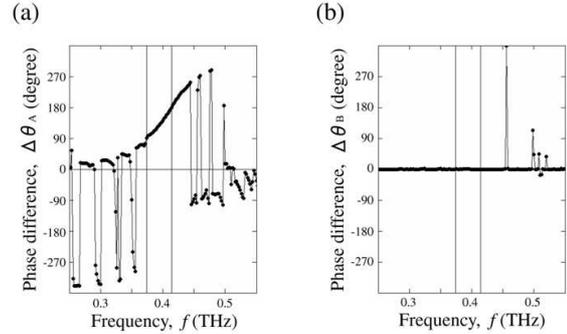}
\caption{\footnotesize (a) $\Delta\theta_{A}$ is the phase-difference, $\theta(11)-\theta(1\overline{1})$.  (b) $\Delta\theta_{B}$ is $\theta(10)-\theta(01)$.
Two vertical solid lines at 0.374 THz and 0.414 THz are ones where the phase-difference is
nearly equal to 90 degrees and 180 degrees, respectively. }
\label{fig:phaseB}
\end{figure}
\begin{figure}[h]
\centering
\includegraphics[width=8.0cm, trim=0  0 0 0]{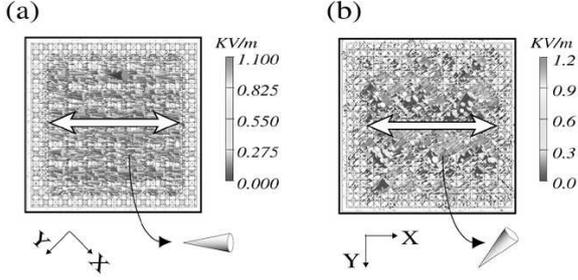}
\caption{\footnotesize Electric-field vectors on the bottom surface (Z=0) at 0.414THz.
The white heavy arrow is the orientation of electric-field vector of the incident wave.
(a) model(1$\overline{1}$). (b) model(10).}
\label{fig:EEfield}
\end{figure}
\hspace*{5mm}Figs.~\ref{fig:EEfield}(a) and \ref{fig:EEfield}(b) show the electric-field vectors in model(1$\overline{1}$) and model(10), respectively. The electric-field vectors are illustrated on the bottom surface (Z=0) at 0.414THz. They are illustrated as cones, which are compound electric-field vectors of the incident waves and reflected ones in the FEM software used. Input power is 1.0 (W).\\ 
\hspace*{5mm}The directions of the cones periodically change with time. 
In figs.~\ref{fig:EEfield}(a) and \ref{fig:EEfield}(b), the time is zero, which corresponds to phase zero. The amplitude of sum of the cones (electric-field vectors) is able to be defined.\\
\hspace*{5mm}The orientation of the incident wave, white heavy arrow in fig.~\ref{fig:EEfield}(a) are [Y~=~$-$X], and the orientation of the cones in X-Y plane are almost entirely parallel to that of the incident wave.\\
\hspace*{5mm}A parallelism tolerance was estimated as follows though the cones are compound electric-field vectors of the incident waves and reflected ones.\\
\hspace*{5mm}Port, $(150\times150) (\mu$m)$^2$  is segmented into $15\times15$ parts and the amplitude of sum of $16\times16$ compound electric-field vectors are calculated in each direction of [Y = X] and [Y~=~$-$X]. Two kinds of amplitude is defined as $E_\mathrm{I+R}(11)$ and $E_\mathrm{I+R}(1\overline{1})$, respectively. \\
\hspace*{5mm}When the orientation of electric-field of the incident wave is PI(1$\overline{1}$), [Y = $-$X] in model(1$\overline{1}$),\\
\hspace*{10mm}$E_\mathrm{I+R}(11)$/$E_\mathrm{I+R}(1\overline{1})=0.0068$  
\\
When the orientation of electric-field of the incident wave is PI(11), [Y = X] in model(11),\\
\hspace*{10mm}$E_\mathrm{I+R}(1\overline{1})$/$E_\mathrm{I+R}(11)=0.0021$  \\
The ratio of components was also estimated at 0.374 THz.\\ 
\hspace*{5mm}$E_\mathrm{I+R}(11)$/$E_\mathrm{I+R}(1\overline{1})=0.0040$: model(1$\overline{1}$) \\
\hspace*{5mm}$E_\mathrm{I+R}(1\overline{1})$/$E_\mathrm{I+R}(11)=0.035$: model(11)  \\
The values of four ratios are very small. 
\hspace*{5mm}In addition, four kinds of reflectivity within PBG is large, which do not have an influence on the digits of above four values, largely.\\
\hspace*{5mm}In experimental aspect, the measured compound (S-p + P-p) reflected-spectra and transmission ones are shown in Fig.A.~\ref{fig:RT10RT11}. The transmittance is very small within BPG.
Fig.A.~\ref{fig:RSRP} shows the measured and separated reflected-spectra, $RS(f)$ and $RP(f)$$^{[2]}$. $\theta$ = 0 degrees and 90 degrees correspond to model(11) and model(1$\overline{1}$), respectively. 
It can be seen that both reflected spectra primarily consist of S-p ones, and they do not almost include P-p ones.
\\
\hspace*{5mm}It is estimated that, in model(11) and model(1$\overline{1}$), the orientations of sum of electric-field vectors of the reflected wave are almost entirely parallel to that of the incident wave at 0.414THz and 0.374THz within PBG.\\ 
\hspace*{5mm}The orientation of the white heavy arrow in fig.~\ref{fig:EEfield}(b) corresponds to PI(10), [X]. The orientation of the cones is an oblique direction. It suggests that, in model(10) and in model(01), the polarization orientation of the reflected wave is extremely different from that of the incident
wave.\\
\section*{Discussions}
\hspace*{5mm}Table~\ref{table:table} shows four kinds of reflectivity, R(XY) in fig.~\ref{fig:RT4} and phase-differences,  $\Delta\theta_{A}$ and  $\Delta\theta_{B}$ in fig.~\ref{fig:phaseB} at 0.374 THz and 0.414THz, respectively.\\
\hspace*{5mm}At each frequency, R(10) and R(01) are nearly equal, and  $\Delta\theta_{B}$ is very small. Two directions, [X] and [Y] are equivalent with polarization features at least within PBG.\\
\hspace*{5mm}On the other hand, at each frequency, R(1$\overline{1}$) and R(11) are not almost the same values, and  $\Delta\theta_{A}$ is very large. Two directions, [Y~=~X] and [Y = $-$X] that are identical in appearance, are not equivalent with polarization features at least within PBG.
These features have already been shown in figs.~\ref{fig:RT4} and ~\ref{fig:phaseA}.\\
\hspace*{5mm}From simulation results, the orientation of electric-field of the reflected wave
is almost parallel to that of the incident wave at 0.374THz and 0.415THz within PBG in the case of model(11) and model(1$\overline{1}$). 
By using the reflectivity and phase-difference of these two models in table~\ref{table:table} and the synthesis of electric-field vectors, the polarization features of the reflected wave in model(10) and model(01) are able to be obtained.\\
\hspace*{5mm}The electric-field vector of the reflected wave in X-Y plane, $E_\mathrm{R}(X, Y)$ is,\\
\hspace*{8mm}$E_\mathrm{R}(X, Y)=E_\mathrm{R0}\times cos[2\pi ft+\delta]$\\
where, $E_\mathrm{R0}$:~amplitude, $f$:~frequency, $t$:~time, $\delta$:~phase-difference.\\
$E_\mathrm{R0}$ is $E_\mathrm{R0}$(XY). The square of $E_\mathrm{R0}$(XY) is proportional to reflectivity, R(XY). \\
\hspace*{5mm}The values of model(11) and model(1$\overline{1}$) in table~\ref{table:table}
are substituted for above expression, and then, attending the orientation of the coordinate axes, the orbits of electric-field of the reflected wave in model(10) at 0.414THz and 0.374THz are able to be obtained.\\
\begin{table}[t]
\centering
\caption{\footnotesize Reflectivity and phase-difference at two frequencies
within PBG in four models}
\includegraphics[width=7.7cm, trim=0  0 0 0]{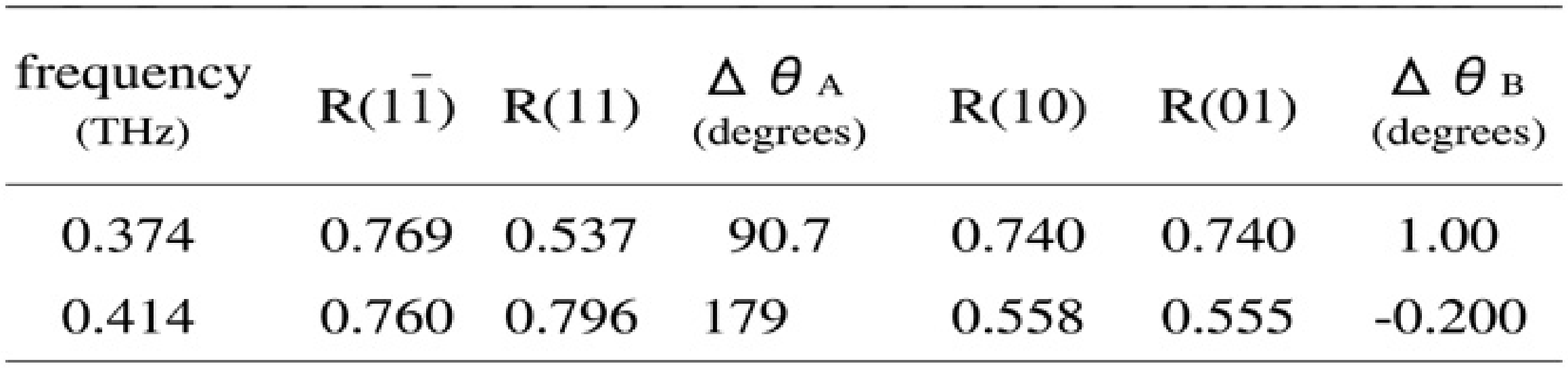}
\label{table:table}
\end{table}
\begin{figure}[t]
\centering
\includegraphics[width=8.0cm, trim=0  0 0 0]{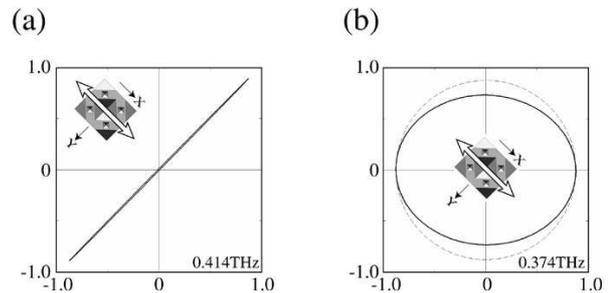}
\caption{\footnotesize Orbit of electric-field vector of the reflected wave. The white heavy arrow is the orientation of electric-field vector of the incident wave.
(a) at 0.414THz. (b)  at 0.374THz. The dashed line is true circle.}
\label{fig:polarization}
\end{figure}
\hspace*{5mm}Fig.~\ref{fig:polarization}(a) shows that the orientation of electric-field of the reflected wave is almost parallel to Y-axis when that of the incident wave is parallel to X-axis in model(10). In other words, the polarization orientation of the reflected wave rotates by 90 degrees for that of the incident wave in model(10). In the same way, in model(01), that of the reflected wave is almost parallel to X-axis when that of the incident wave is parallel to Y-axis.\\
\hspace*{5mm}Fig.~\ref{fig:polarization}(b) shows the reflected wave is near-circularly-polarized wave at 0.374 THz in model(10) and model(01). The orientation of long axis and short one of the orbit is [Y = $-$X] and [Y = X], respectively.\\
\hspace*{5mm}On the other hand, within BGX, the difference of size of the measured reflected spectra, RS(11) and RS(1$\overline{1}$) is much less than that of the simulation spectra as can be seen from fig.A.~\ref{fig:RSRP}, and then RP(11) and RP(1$\overline{1}$) are very small\footnote{The diameter of spot size of the incident wave in measurement is about 6 $mm$ (20a), which is much larger than the FEM port size, $5a\times5a$. 
It seems that it is one of differences between measured spectra and FEM ones.}.
It is expected that the orbit of the measured reflected spectra is further circular one at 0.374 THz.\\
\hspace*{5mm}The phase-difference above does not probably occur in general when the wavelength of the incident wave is much larger than the PC's lattice constant.\\
\section*{Conclusions}
\hspace*{5mm}The analyses of phase, size and orientation of electric-field of the reflected wave indicated that two directions, [Y = $\pm$X] identical in appearance are not identical for
polarization anisotropy in the 3D-PC of this work, and then revealed the reason why the polarization orientation of the reflected wave is rotated by 90 degrees at 0.414THz within PBG when that of the incident wave is parallel to X-axis (or Y-axis). In addition, it was found that the orbit of electric-field of the reflected wave at 0.374THz within PBG becomes near-circularly-polarized one for the orientation of electric-field of above incident wave.\\ 

\newpage
\section*{Appendix}
\setcounter{figure}{0}
\renewcommand{\figurename}{Fig.~A.}
\begin{figure}[h]
\centering
\includegraphics[width=6.5cm, trim=0 0 0 0]{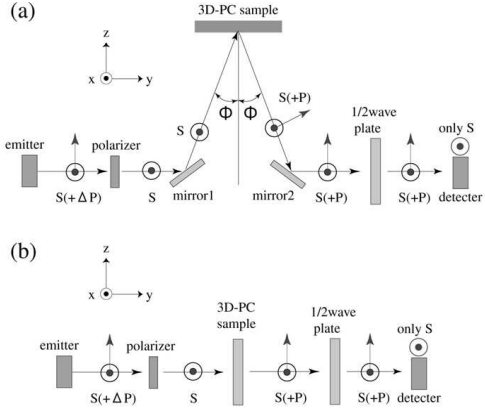}
\caption{\footnotesize Schematic diagram of the measurement system with a THz-TDS. (a) reflection measurement system (b) transmission measurement system. The polarization orientation of a THz wave from the emitter is S-p ∥ x-axis and the detector detects only S-p. Another polarization perpendicular to S-p is P-polarization (P-p) that is included in the incidence plane $\|$ z-axis. The designed 1/2 wave plate converts S-p into P-p and P-p into S-p at around 0.42 THz. This plate was used in the case of P-p detection. S(+P) means S-p, P-p or the mixing of S-p and P-p. In fig.~A\ref{fig:RTsystem}(a), the 3D-PC sample is so horizontally set that the layered direction is parallel to z-axis, that is, Z-axis in 
fig.~\ref{fig:latticeA} is parallel to z-axis. In this work, the 3D-PC sample was rotated in plane (001) instead of the S-p incident wave, relatively. The incident angle, $\phi$ is $7^{\ \circ}$, which is the perpendicular incidence approximately.}
\label{fig:RTsystem}
\end{figure}
\begin{figure}[t]
\centering
\includegraphics[width=7.4cm, trim=10 0 0 220]{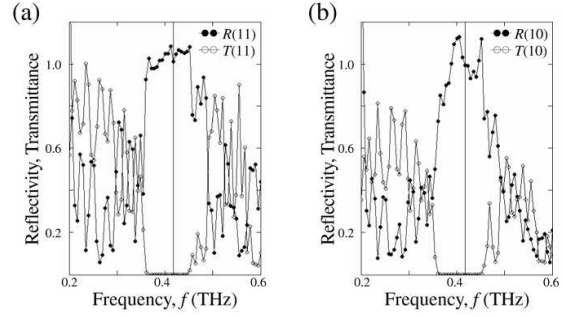}
\caption{\footnotesize (a) Measured compound (S-p + P-p) reflected spectra (closed circle), $R$(11) and transmission ones (open circle), $T$(11). (b) Similarly, reflected spectra, $R$(10) and transmission ones, $T$(10). The vertical solid line is 0.42 THz-line.}
\label{fig:RT10RT11}
\end{figure}
\begin{figure}[h]
\centering
\includegraphics[width=7.0cm, trim=10 0 0 220]{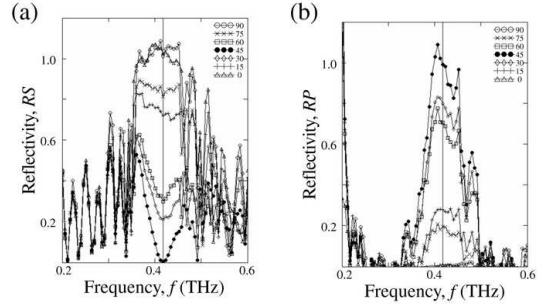}
\caption{\footnotesize Measured reflected spectra, (a) $RS(f)$, (b) $RP(f)$ for sample's
rotation angles $\theta$ from 0 degrees to 90 degrees per 15 degrees.  
The electric-field direction of incident wave of $\theta$ = 0 degrees, 45 degrees, and 90 degrees corresponds to PI(11), PI(10) and PI(1$\overline{1}$) in fig.~\ref{fig:IXY}, respectively. The vertical solid line is 0.42 THz-line. $RS(f)$ is the S-p reflectivity whose polarization orientation is parallel to that of the incident wave. $RP(f)$ is the P-p reflectivity whose polarization orientation is perpendicular to S-p.}
\label{fig:RSRP}
\end{figure}

\end{document}